\documentclass{iau} 

\usepackage{graphicx}
\usepackage{xcolor}
\usepackage{multicol,graphicx, wrapfig,xspace,float}
\usepackage[hyphens]{url}
\usepackage{hyperref}
\hypersetup{
     colorlinks  = true,
     linkcolor   = blue, 
     anchorcolor = BrickRed,
     citecolor   = blue,
     filecolor   = blue,
     menucolor   = blue,
     runcolor    = cyan,
     urlcolor    = blue
}
\usepackage[authoryear]{natbib}

\newcommand{\lcdm}[0]{$\Lambda$CDM\xspace}

\pubyear{2019}
\volume{355} 
\jname{The Realm of the Low-Surface-Brightness Universe}
\editors{D. Valls-Gabaud, I. Trujillo \& S. Okamoto, eds.}

\title[Tidal Structures In Modified Gravity] 
{Origin Of Tidal Structures In Modified Gravity}

\author[Michal B\'ilek et al.]   
{Michal B\'ilek$^{1,*}$, Ingo Thies$^{2}$, Pavel Kroupa$^{2,3}$, Benoit Famaey$^{1}$}

\affiliation{$^1$ Universit\'e de Strasbourg, CNRS, Observatoire astronomique de Strasbourg (ObAS), UMR 7550, 67000 Strasbourg, France; $^2$ Helmholtz-Institut f\"ur Strahlen- und Kernphysik, Nussallee 14-16, 53115 Bonn, Germany; $^3$ Charles University in Prague, Faculty of Mathematics and Physics, Astronomical Institute, V Hole\v sovi\v ck\' ach 2, 180 00 Praha 8, Czech Republic \\$^*$ email: {\tt bilek@astro.unistra.fr}}

\begin{document}

\maketitle

\begin{abstract}
The missing mass problem has not been solved decisively yet. Observations show that if gravity is to be modified, then the MOND theory is its excellent approximation on galactic scales. MOND suggests an adjustments of the laws of physics in the limit of low accelerations. Comparative simulations of interacting galaxies in MOND and Newtonian gravity with dark matter revealed two principal differences: 1) galaxies can have close flybys without ending in mergers in MOND because of weaker dynamical friction, and 2) tidal dwarf galaxies form very easily in MOND. When this is combined with the fact that many interacting galaxies are observed at high redshift, we obtain a new perspective on tidal features: they are often formed by non-merging encounters and tidal disruptions of tidal dwarf galaxies. Here we present the results from our self-consistent MOND $N$-body simulation of a close flyby of two galaxies similar to the Milky Way. It turns out that most types of the structures that are traditionally assigned to galaxy mergers can be formed by non-merging encounters, including tidal arms, bridges, streams, shells, disk warps, thick disks, and most probably also disks of satellites. The success of MOND in explaining the dynamics of galaxies hints us that this way of formation of tidal structures should be considered seriously.

\end{abstract}

\firstsection 

\section{Motivation}


Physics faces two big challenges: the missing mass problem and the absence of a successful quantum gravity theory. Perhaps the two questions are related and gravity does not behave in the standard way if the curvature of spacetime is small. This is actually hinted by observations since rotation curves\footnote{A rotation curve is the rotation velocity as a function of galactocentric radius.} of spiral galaxies are fully determined by the distribution of baryonic matter (e.g., \citealp{mcgaugh16}) following the MOND modified gravity theory \citep{milg83a, famaey12}. MOND can fit accurately rotation curves of galaxies spanning a wide range of masses, sizes and gas fractions (e.g., \citealp{gentile11,mcgaugh16}). The predictability of rotation curves from the distribution of baryons is hard to explain if the rotation curves are largely determined by dark matter, as the standard Newtonian theory assumes. Testing MOND in dwarf and elliptical galaxies is observationally difficult but the existing results are encouraging (e.g., \citealp{anddwarfii,lelli17,bil19}). Weak gravitational lensing \citep{milg13} and dynamics of galaxy groups \citep{milg19} indicate that MOND works well also on the scales of hundreds of kpc in sparse environments. The validity of MOND remains unclear for galaxy clusters and cosmology \citep{milgschol}. 

\section{Galaxy interactions in MOND}

The above facts lead us to conclude that if gravity is to be modified, then it has to be very similar to MOND on the scales of galaxies and galaxy groups. It then makes sense to  apply MOND on galaxy interactions. Comparative simulations of galaxy interactions in MOND and in the Newtonian gravity with dark matter ($\Lambda$CDM) identified two main differences \citep{nipoti07,tiret07c,combtir10}. 1) Dynamical friction\footnote{Dynamical friction is a process that transforms the orbital energy and momentum of the two galaxies into the internal energy and momentum of their constituents, i.e. stars and dark matter particles. It creates a force acting against the direction of motion of the galaxies and causes the galaxies to merge.} during interactions of comparable galaxies is weaker in MOND and therefore galaxies can encounter relatively closely without merging within the age of the Universe. Many observed interacting pairs that are traditionally called ``mergers'' can actually be non-merging galaxy flybys as emphasized by \citet{kroupacjp}. This is a good news since $\Lambda$CDM simulations suffer of overproducing of galactic bulges because of  frequent mergers (\citealp{kroupacjp}, see also \citealp{combes14}). 2) Tidal dwarf galaxies, i.e. small galaxies forming in tidal tails of interacting galaxies, form much easier in MOND. The observed organization of satellites into disks, that is difficult to explain with the $\Lambda$CDM theory, gets explained easily if the satellites are tidal dwarf galaxies \citep{kroupacjp}.

\section{Tidal structures without mergers}


Observations teach us that galaxy interactions become more frequent with increasing redshift. Approximately 20\% of galaxies with stellar masses over $10^{9.5}$\,M$_\odot$ are seen interacting at the redshift of 1 (see, e.g., the compilation in Fig.~6 of \citealp{bridge10}). In addition, the typical velocity difference of neighboring galaxies in the field is only 30\,km\,s$^{-1}$ \citep{karachentsev12}, favoring close encounters. We can then expect assuming modified gravity that many tidal structures were created by non-merging galaxy flybys. Further tidal features were formed by eroding tidal dwarf galaxies by tidal forces. This is different from the classical view that tidal features are remnants of mergers of primordial galaxies. This brings up the question of what types of tidal structures can be formed by the non-merging encounters.

We witnessed the formation of structures that are traditionally associated with mergers in our self-consistent simulation of the history of the Local Group in MOND \citep{bil18}. The initial conditions were set so that the simulated Milky Way and M\,31 reproduced their observed masses, sizes, disk inclinations, distance, and relative velocity (nothing else was tuned). It had been known before from analytic calculations that MOND implies that these galaxies had a close encounter at the redshift of 1-2 \citep{zhao13}. Our simulation showed that the galaxies reached a distance of 24\,kpc at the pericenter. The encounter induced the formation of long tidal tails in the Milky Way one of which was captured by M\,31.  Such an encounter  is very unlikely in $\Lambda$CDM since the galaxies would merge in 1-2\,Gyr after the pericenter. In contrast, the galaxies in the MOND simulation were not even close to merging at the end of the simulation 6.5\,Gyr from today. 

Here follows a parade of tidal structures that were formed during the course of our simulation of this non-merging encounter.
\begin{figure}[H]
	\centering
	\begin{minipage}[b]{0.4\linewidth}
	\centering
	\textbf{Tidal tails}:
  \includegraphics[width=1\textwidth]{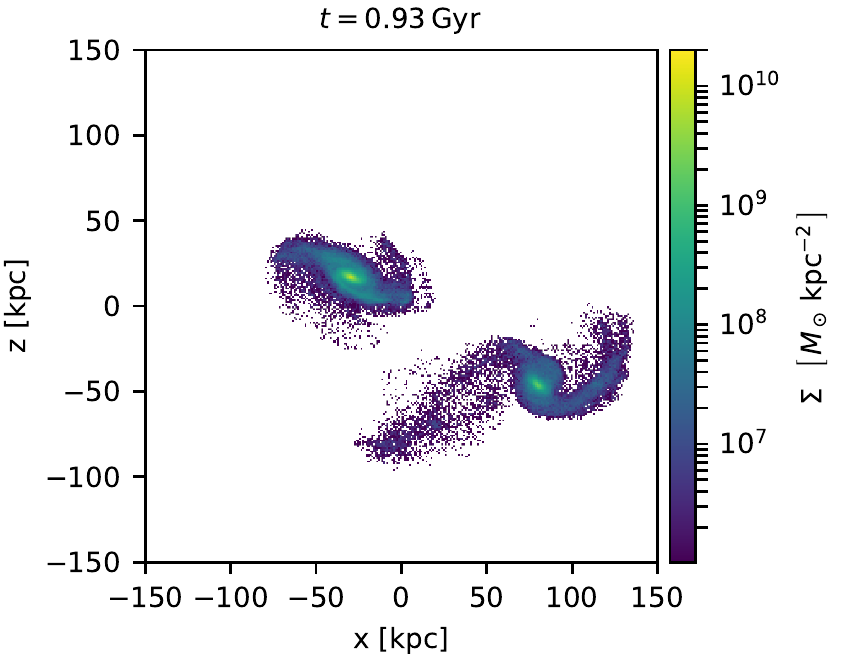}
\end{minipage}
\hspace{3em}
\begin{minipage}[b]{0.4\linewidth}
\centering
\textbf{Tidal bridge}:
  \includegraphics[width=1\textwidth]{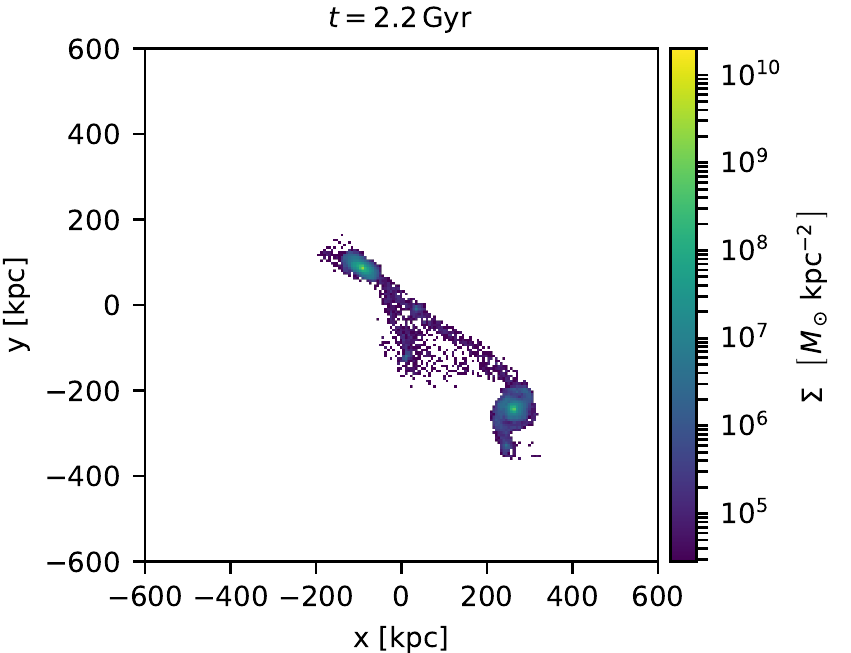}
\end{minipage}
\end{figure}

\textbf{Tidal streams} around the simulated M\,31 (left) reproduce the loop-like morphology and size of the real streams (right, \citealp{ferguson16}). The simulated streams host tidal dwarf galaxies.
\vspace{-1cm}
\begin{figure}[H]
	\centering
	\begin{minipage}[b]{0.43\linewidth}
	\centering
  \includegraphics[width=1\textwidth]{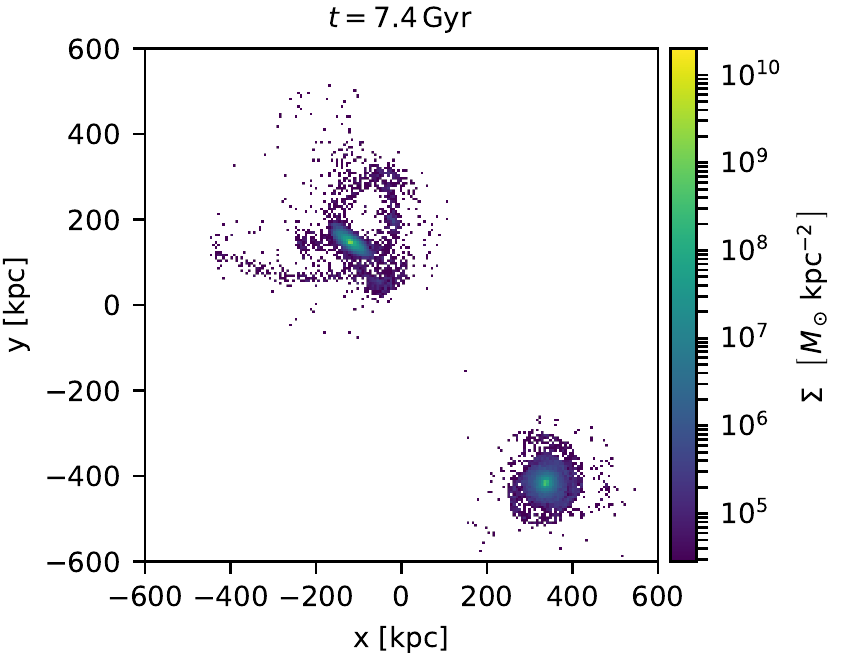}
\end{minipage}
\hspace{3em}
\begin{minipage}[b]{0.35\linewidth}
\centering
  \includegraphics[width=1\textwidth]{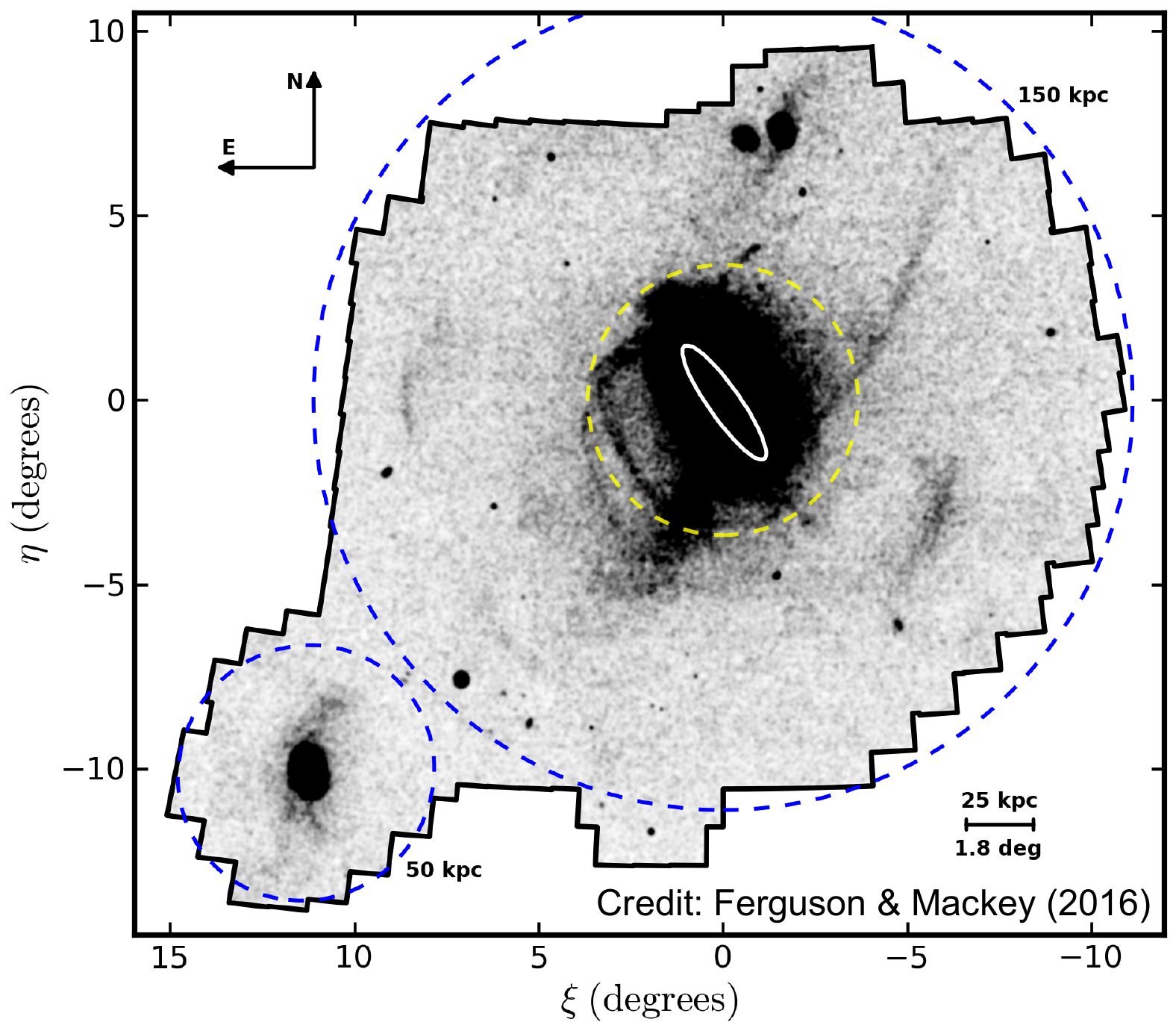}
\end{minipage}
\end{figure}

The encounter caused a \textbf{disk warp} in the simulated Milky Way (left) with a magnitude and orientation similar to the real one (right, \citealp{levine06}).
\vspace{-5ex}
\begin{figure}[H]
	\centering
	\begin{minipage}[b]{0.4\linewidth}
	\centering
  \includegraphics[width=1\textwidth]{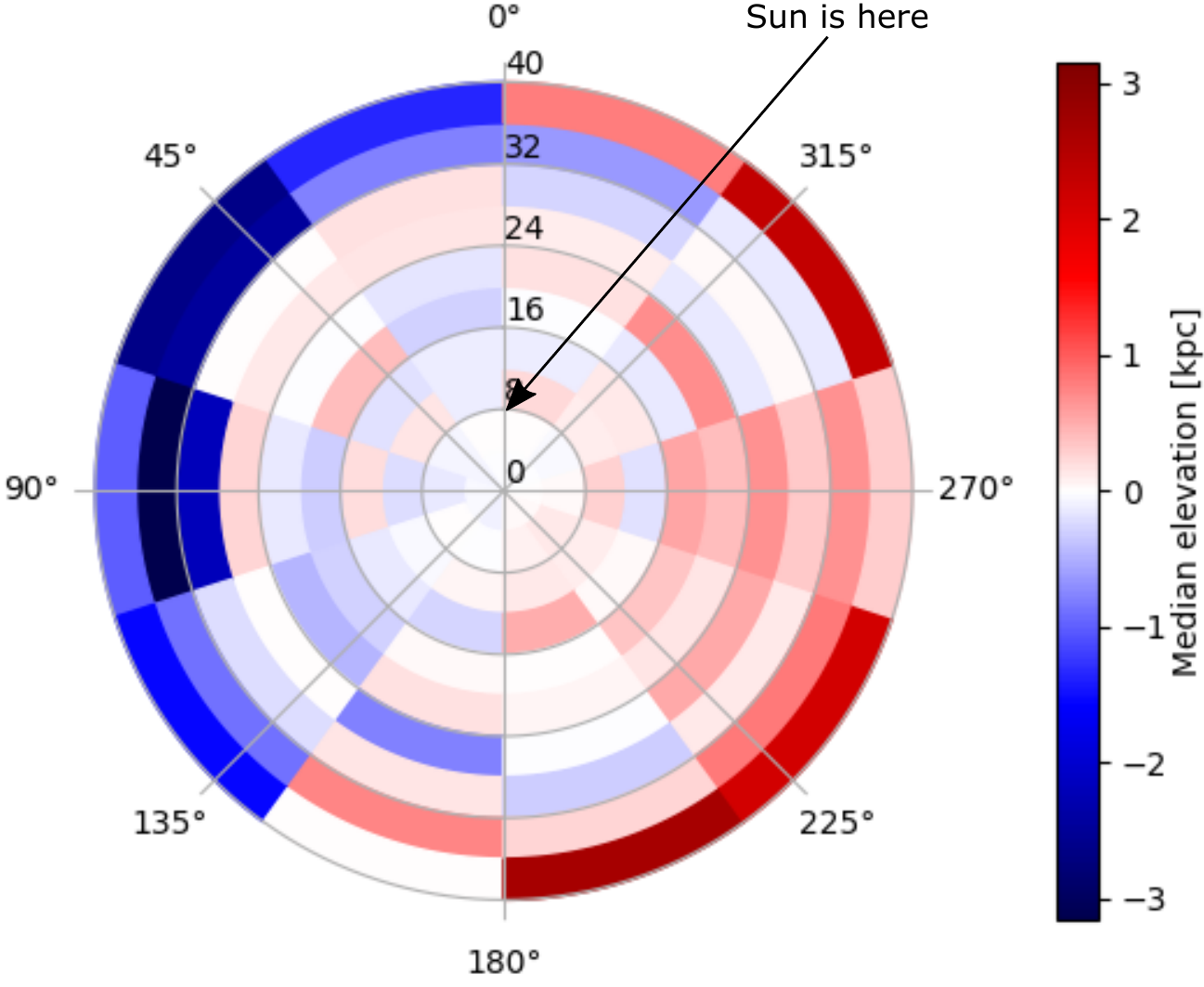}
\end{minipage}
\hspace{3em}
\begin{minipage}[b]{0.4\linewidth}
\centering
  \includegraphics[width=1\textwidth]{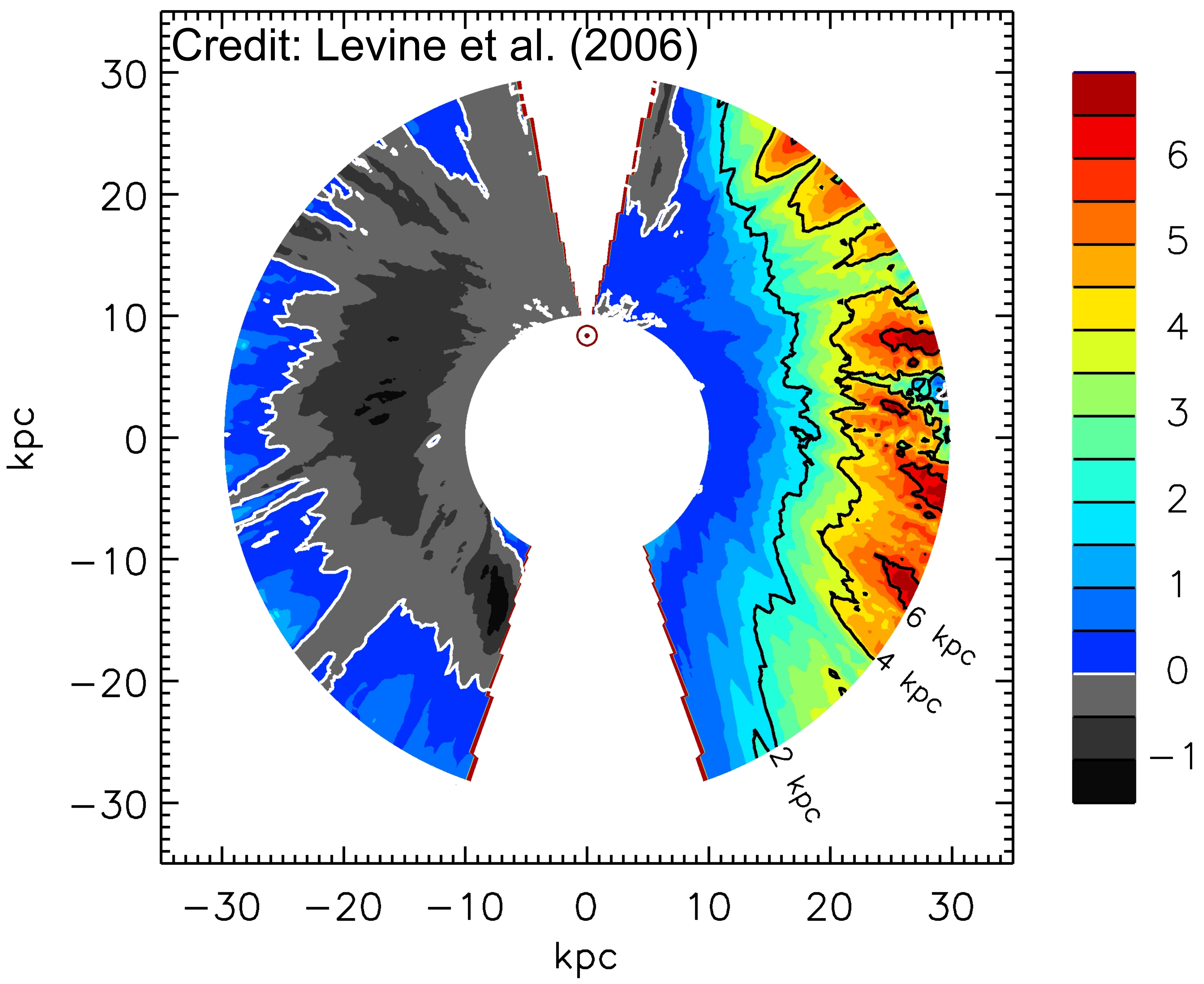}
\end{minipage}
\end{figure}

The simulation suggested the origin of the \textbf{thick disks} (left).	\textbf{Shells} were not formed in our simulation but \citet{hq88} demonstrated that non-merging flybys can produce them (right).
\begin{figure}[H]
	\centering
	\begin{minipage}[b]{0.4\linewidth}
	\centering
  \includegraphics[width=1\textwidth]{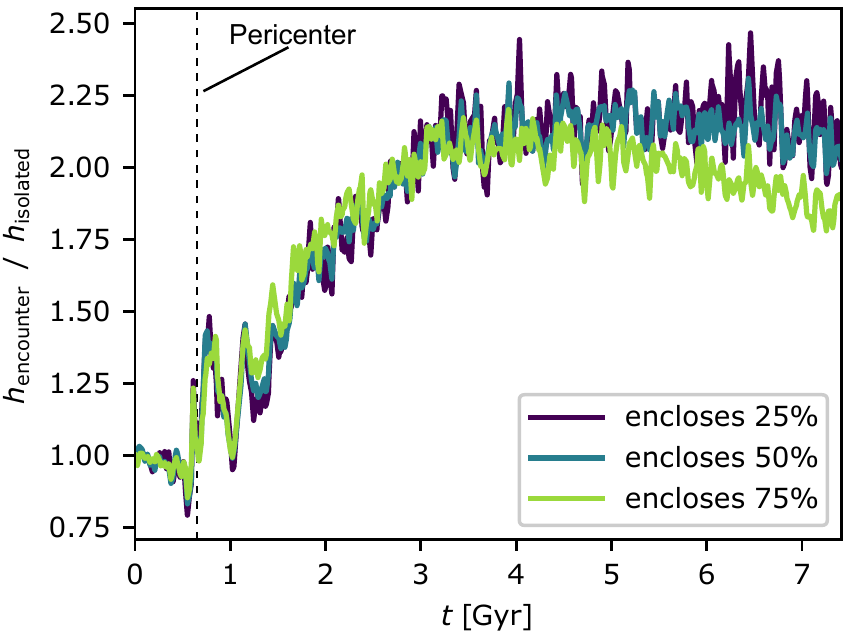}
\end{minipage}
\hspace{3em}
\begin{minipage}[b]{0.3\linewidth}
\centering
  \includegraphics[width=1\textwidth]{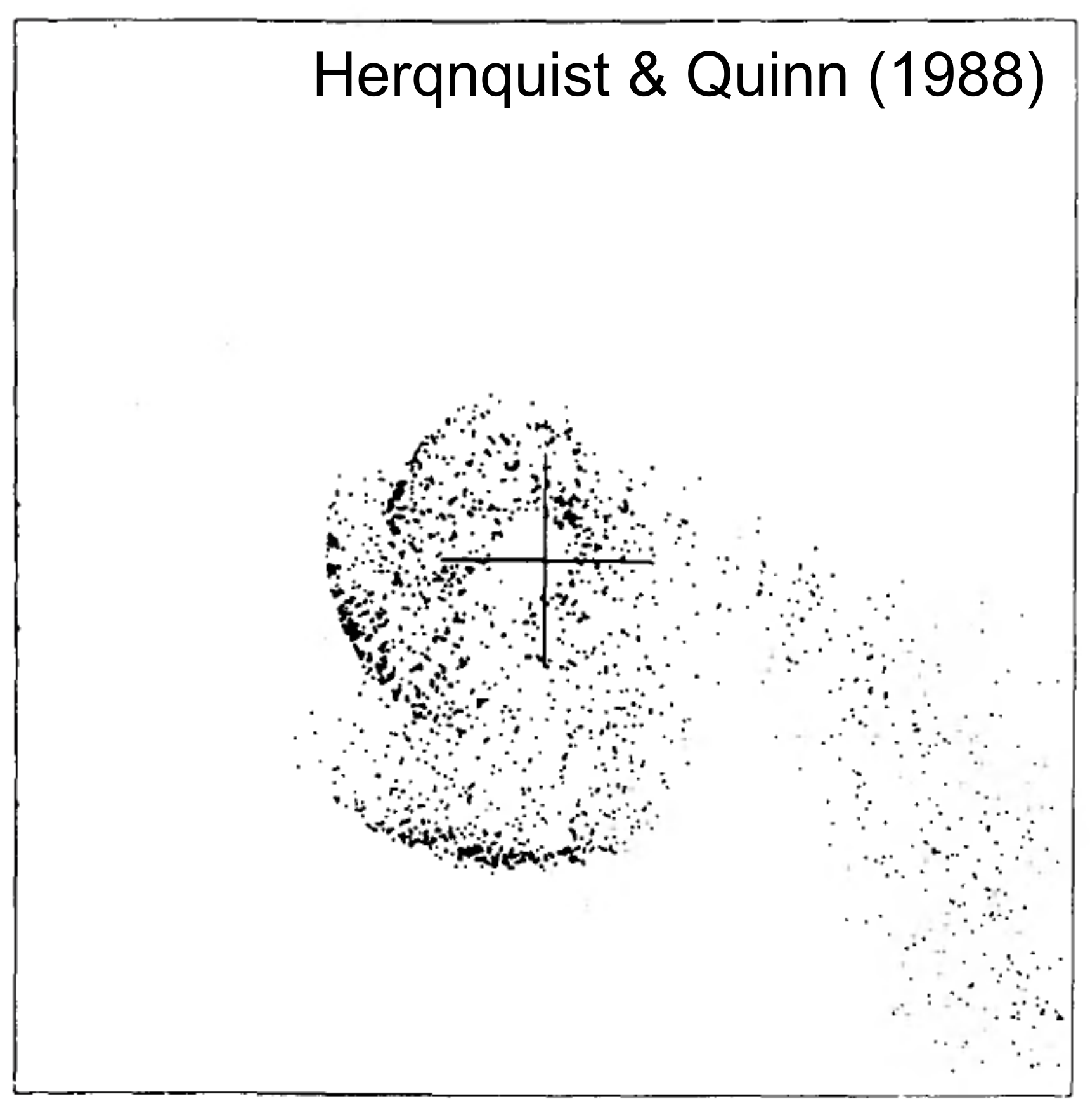}
\end{minipage}
\end{figure}

The structure at the simulated M\,31 (left) can also be identified with the observed \textbf{disk of satellites} (right) (similar size, mass, rotation, orientation). The color indicates the line-of-sight velocity from the position of the Sun.
\begin{figure}[H]
	\centering
	\begin{minipage}[b]{0.42\linewidth}
	\centering
  \includegraphics[width=1\textwidth]{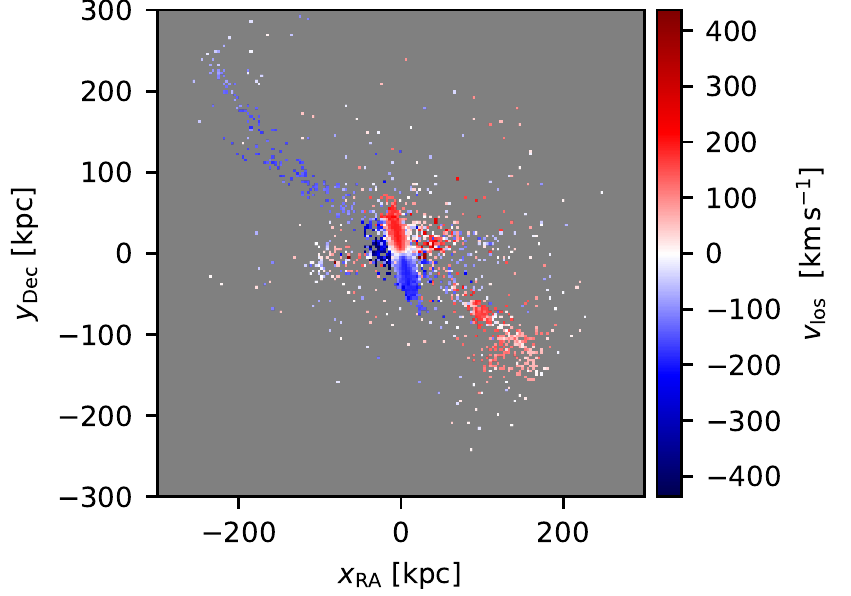}
\end{minipage}
\hspace{3em}
\begin{minipage}[b]{0.32\linewidth}
\centering
  \includegraphics[width=1\textwidth]{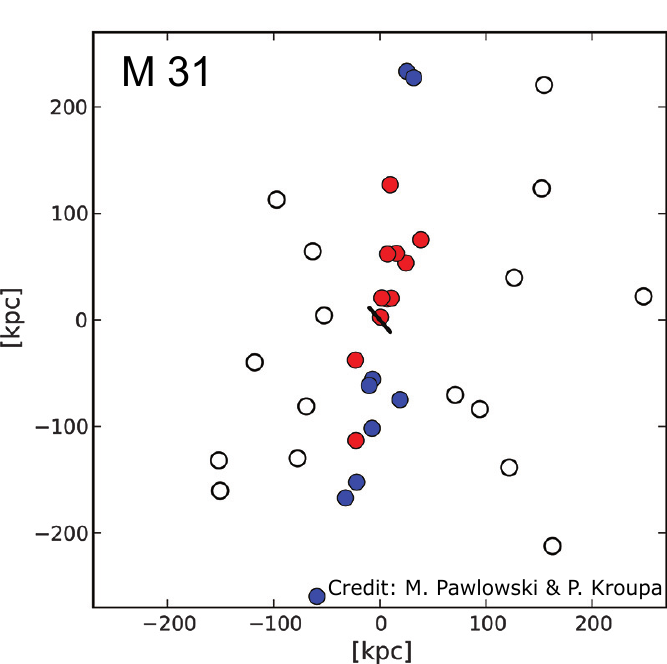}
\end{minipage}
\end{figure}

\section{Role of galaxy mergers in MOND}
Mergers of comparable galaxies can, of course, happen even in MOND if the galaxies appear close to each other  with a low relative velocity or if the stellar bodies of the galaxies hit each other \citep{combtir10}.  Interestingly, the results of \citet{ciotti04} and \citet{nipoti08} suggest that dynamical friction in MOND becomes stronger than in \lcdm for interacting galaxies whose masses are very different (minor mergers). Real simulations of such interactions are however still missing.

\begin{multicols}{2}
\bibliographystyle{aa-short}
\bibliography{citace}
\end{multicols}

\end{document}